\documentstyle[11pt,aaspp4]{article}

\def\lea{\mathrel{<\kern-1.0em\lower0.9ex\hbox{$\sim$}}}
\def\gea{\mathrel{>\kern-1.0em\lower0.9ex\hbox{$\sim$}}}
\slugcomment{Accepted by The Astrophysical Journal Letters}

\lefthead{Sarajedini et al.}
\righthead{M33 Globular Clusters}

\begin{document}

%%AJ:
%%\def\et{\hbox{\em{et~al.\ }}}
%%ApJL:
\def\et{\hbox{et~al.\ }}

\def\feh{\hbox{[Fe/H]}}
\def\ofe{\hbox{[O/Fe]}}
\def\deg{\hbox{$^\circ$}}
\def\sun{\hbox{$\odot$}}
\def\earth{\hbox{$\oplus$}}
\def\lesssim{\mathrel{\hbox{\rlap{\hbox{\lower4pt\hbox{$\sim$}}}\hbox{$<$}}}}
\def\gtrsim{\mathrel{\hbox{\rlap{\hbox{\lower4pt\hbox{$\sim$}}}\hbox{$>$}}}}
\def\sq{\hbox{\rlap{$\sqcap$}$\sqcup$}}
\def\arcmin{\hbox{$^\prime$}}
\def\arcsec{\hbox{$^{\prime\prime}$}}
\def\fd{\hbox{$.\!\!^{\rm d}$}}
\def\fh{\hbox{$.\!\!^{\rm h}$}}
\def\fm{\hbox{$.\!\!^{\rm m}$}}
\def\fs{\hbox{$.\!\!^{\rm s}$}}
\def\fdg{\hbox{$.\!\!^\circ$}}
\def\farcm{\hbox{$.\mkern-4mu^\prime$}}
\def\farcs{\hbox{$.\!\!^{\prime\prime}$}}
\def\fp{\hbox{$.\!\!^{\scriptscriptstyle\rm p}$}}
\def\micron{\hbox{$\mu$m}}
 
\let\la=\lesssim
\let\ga=\gtrsim

\title{HST WFPC2 Color-Magnitude  Diagrams of Halo\\ 
Globular Clusters in M33: Implications for the \\ Early Formation History
of the Local Group\altaffilmark{1}}

\author{Ata Sarajedini\altaffilmark{2,}\altaffilmark{3}, 
Doug Geisler\altaffilmark{4},
Paul Harding\altaffilmark{5}, and
Robert Schommer\altaffilmark{6}}
%\affil{San Francisco State Univerity, Department of Physics and Astronomy, \\ 
%1600 Holloway Avenue, San Francisco, CA  94132 \\ {\it ata@stars.sfsu.edu}}

%\author{Doug Geisler}
%\affil{Kitt Peak National Observatory,
%National Optical Astronomy Observatories,\\
% P. O. Box 26732, Tucson, AZ  85726 \\ {\it dgeisler@noao.edu}}
 
%\author{Paul Harding}
%\affil{Steward Observatory, University of Arizona, Tucson, AZ 85721 \\ 
%{\it harding@as.arizona.edu}}

%\author{Robert Schommer}
%\affil{Cerro Tololo Inter-American Observatory,
%National Optical Astronomy Observatories,\\
%Casilla 603, La Serena, Chile \\ {\it rschommer@noao.edu}}

\altaffiltext{1}{Based on observations made with the NASA/ESA Hubble
Space Telescope obtained at the Space Telescope Science Institute,
which is operated by the Association of Universities for Research in
Astronomy, Incorporated, under NASA contract NAS5-26555.}
\altaffiltext{2}{Hubble Fellow}
\altaffiltext{3}{San Francisco State Univerity, Department of Physics and Astronomy, 1600 Holloway Avenue, San Francisco, CA  94132, {\it ata@stars.sfsu.edu}}
\altaffiltext{4}{Kitt Peak National Observatory,
National Optical Astronomy Observatories,
P. O. Box 26732, Tucson, AZ  85726, {\it dgeisler@noao.edu}}
\altaffiltext{5}{Steward Observatory, University of Arizona, 
Tucson, AZ 85721, {\it harding@as.arizona.edu}}
\altaffiltext{6}{Cerro Tololo Inter-American Observatory,
National Optical Astronomy Observatories,
Casilla 603, La Serena, Chile, {\it rschommer@noao.edu}}

% The abstract environment prints out the receipt and acceptance dates
% if they are relevant for the journal style.  For the aasms style, they
% will print out as horizontal rules for the editorial staff to type
% on, so long as the author does not include \received and \accepted
% commands.  This should not be done, since \received and \accepted dates
% are not known to the author.

\begin{abstract}

We have constructed color-magnitude diagrams (CMDs) for ten globular 
clusters in the
halo of the nearby spiral galaxy M33 based on Hubble
Space Telescope Wide-Field Planetary Camera 2 observations in the
F555W ($\sim$V) and F814W ($\sim$I) filters. 
%The cluster photometry extends
%$\sim$1.5 magnitudes below the horizontal branch (HB). 
These data
reveal the morphology of the HB and allow us to
estimate the cluster metallicity using the shape and color of the red giant
branch. The principal result we report herein is that 8 of
the 10 clusters possess exclusively red HB morphologies yet their
metallicities are as metal-poor as $[Fe/H] = -1.6$. 
Indeed, these 8 clusters present basically only giant branch clumps 
reminiscent of
intermediate age star clusters in the Magellanic Clouds. In addition,
two of the clusters form a second parameter pair which have similar
metallicities but very disparate HB types. Under the assumption that
cluster age is the global second parameter, the average age of
halo globular clusters
in M33 appears to be a few Gyr younger than halo clusters in the Milky
Way. Using the observed properties of 
HB stars in M31 and M33 along with published main-sequence turnoff ages for 
the globular clusters in the Milky Way, LMC, SMC, and the Sagittarius
dwarf spheroidal (Sgr), we attempt to sketch
the early formation history of these galaxies. This indicates that
the Milky Way, M31, M33, the LMC, and Sgr all experienced their first epoch of
cluster formation soon after the Big Bang. Three to four Gyr later, the SMC
began to form its first generation of clusters; the {\it bulk} of the 
M33 clusters 
formed later still. We note that the halo clusters in M33 formed
over a much larger time period than those in the Milky Way and M31.

\end{abstract}

% The different journals have different requirements for keywords.  The
% keywords.apj file, found on aas.org in the pubs/aastex-misc directory, 
% contains a list of keywords used with the ApJ and Letters.  These are 
% usually assigned by the editor, but authors may include them in their 
% manuscripts if they wish. 

%\keywords{globular clusters,peanut clusters,bosons,bozos}

% In the first two sections, you should notice the use of the LaTeX \cite
% command to identify citations.  The citations are tied to the
% reference list via symbolic KEYs.  We have chosen the first three
% characters of the first author's name plus the last two numeral of the
% year of publication.  The corresponding reference has a \bibitem
% command in the reference list below.
%
% Please see the AASTeX manual for a more complete discussion on how to make
% \cite-\bibitem work for you.   

\section{Introduction}

The globular clusters of the Milky Way (MW) probe the earliest formation epochs
of our Galaxy. Identifying similar clusters in other  galaxies allows us
to study the properties of these galaxies soon after their formation. 
In addition to the MW, a number of other Local Group galaxies
contain globular clusters; among these are the Sagittarius dwarf spheroidal
galaxy (Sgr), 
the Large and Small Magellanic Clouds (LMC and SMC), the Fornax dwarf 
spheroidal galaxy, M31, and M33, which is the topic of the present work.
We have constructed CMDs for a sample of halo
globular clusters in M33 in order to estimate their metallicities and
HB morphologies. By comparing these properties with
those of globular clusters in other Local Group galaxies, we hope to
construct a formation chronology for the Local Group. 

%The next section
%describes our sample selection along with a brief description of the
%observations and reductions. We leave the details to a future paper
%(Sarajedini et al. \cite{sarea1998}).
%Section 3 presents our results and Sections 4 and 5 discuss the significance
%of these findings. 

\section{Sample Selection, Observations and Reduction}

We are interested in the oldest star clusters in the halo of M33.
As a result, our target selection was primarily based on the integrated
color and kinematics of the clusters. In particular, we inspected Fig. 2 of Schommer et al. (\cite{schea1991}),
which shows the difference between the cluster velocity and the disk
velocity as a function of the integrated cluster color. This difference is
essentially zero for the blue clusters but begins to show a significant
dispersion redward of $(B-V)\sim0.6$, indicating that clusters
redder than this value are likely to be located in the halo of M33.
We were also interested in sampling as much of the spatial and radial
extent of M33 as possible. All of these points lead us to our target list of
10 M33 clusters whose properties are tabulated in Table 1. 
Assuming $(m-M)_0 = 24.5$ for M33 (van den Bergh \cite{va1991}), these 
clusters have --9$\lea$$M_V$$\lea$--7, placing them squarely among the
Galactic halo globular clusters in luminosity (Harris \cite{ha1997}).
In addition, our sample includes
all clusters deemed by Christian \& Schommer (\cite{chsc1983}) to have ages 
$\geq 4$ Gyr.
Previous studies, including Christian \& Schommer (\cite{chsc1983}), 
Cohen et al. (\cite{copese1984}),
Christian \& Schommer (\cite{chsc1988}), 
Schommer et al. (\cite{schea1991}), and Brodie \& Huchra (\cite{brhu1991}),
indicate that, while M33 possesses clusters with a wide age range, 
the clusters we selected are among the oldest clusters in this galaxy.
To sum up, our target list of 10 M33 clusters are similar to Galactic 
halo globular clusters in terms of their luminosities, colors, kinematics,
measured metallicities, and implied ages as they were known at the time. 

The observations used in the present work were obtained with the Wide-Field
Planetary Camera 2 on HST as part of our cycle 5 program (GO-5914). 
Each cluster was centered in the PC1 chip.
%and the spacecraft was rotated to maximize the scientific return and to
%avoid including bright stars in the field. 
We observed each target for 4 orbits,
yielding a total exposure time of 4800~s in F555W and 5200~s in F814W. 

%; the first orbit began with guide star acquisition after which one 2200s
%CR-SPLIT exposure was taken in the F555W ($\sim$V) filter. In the subsequent 
%three orbits, after guide star reacquisition, one more 
%CR-SPLIT F555W image of 2600s duration was taken as well as two 2600s 
%CR-SPLIT exposures in the F814W ($\sim$I) filter. Between each exposure, 
%the spacecraft was offset by an average of 4.5 PC1 pixels in a 
%cross pattern in order to expedite the forthcoming ALLFRAME reduction 
%(see below). 
%Each of the raw HST/WFPC2 images was reprocessed using the latest
%`recommended' calibration exposures. In all cases, the only difference 
%between this and the pipeline processing was that a more appropriate 
%`dark current' exposure was used. 

After reprocessing using the latest `recommended' calibration exposures,
the program PC1 frames were reduced using the procedure outlined by
Silbermann et al. (\cite{silea1996}), which describes the techniques used 
by the HST `Cepheid Distance Scale' key project. Further details are
provided in Sarajedini et al. (\cite{sarea1998}). In summary, empirical
point spread functions (PSFs), kindly provided by P. Stetson,
 were fitted to the detected stellar profiles on
each CCD frame using the ALLFRAME 
%iterative multi-image crowded-field photometry 
package. The resultant photometry was edited and matched to
form instrumental colors. Aperture corrections derived from our frames
and the photometric transformations of Silbermann et al. (\cite{silea1996}) 
were applied to bring the photometry onto the standard system.
Along with the typical error in the aperture corrections 
of $\sim$0.08 mag, the estimated uncertainty of $\sim$0.05 mag in the 
transformations leads to a total uncertainty of $\sim\pm$0.09 mag in the 
photometric zeropoints.

\section{Results}

A preliminary visual inspection of the cluster CMDs revealed a rather
surprising result. Although we had selected the M33 clusters to be
analogous to Galactic globular clusters in every respect, 
all but two of the M33 halo clusters possess extremely red HBs, or red clumps
lying essentially on top of the cluster RGBs. 
%It is
%probably more correct to refer to these HBs as red clumps. 
%We 
%discuss the quantitative assessment of the metallicities and HB
%morphologies below. For now, 
%We show the 
Color-magnitude diagrams of M33-M9 and M33-C20 are shown in
in Figs. 1a and 1b, respectively. These include stars between 1 and
2.8 arcsec of the cluster centers. The solid line is the fiducial sequence
of the Galactic globular cluster M5 (Sandquist et al. \cite{sandea1996})
which has a metallicity between
that of M9 and C20. In addition, the lower panels of
Fig. 1 show the CMDs of M3 (Buonanno et al. \cite{buoea1994}), 
which has a metallicity similar to that of M33-M9, and NGC 362 
(Harris \cite{har1982}), which is similar to M33-C20 in metallicity. 
Both of these Galactic globulars are classified as `younger halo' clusters
by Zinn (\cite{zinn1993}). Figure 1 is an illustration of the fact
that, while two of our M33 clusters
(M9 and U77) have HB morphologies similar to the `younger-halo' MW 
clusters, {\bf the majority of the M33 halo clusters (8 of the 10 in
our sample) possess HBs that are significantly redder than any Milky Way 
globular clusters at their metallicity}. This is the primary result of 
the present work.

%Figure 1 indicates that the oldest
%globular clusters in M33 represented by the two with both blue and red
%HBs (U77 and M9) appear to be
%the same age as the younger-halo globulars in the MW; in
%contradistinction, the majority of globular clusters in M33,
%represented by the remaining 8 clusters, all of which have completely
%red HBs,   at the same 
%metallicity, the red HB globular clusters in M33 (8 of the 10 in our sample)
%halo globular    
%clusters in M33 that possess red HB morphologies than analogous 
%clusters in the MW, even those that are amongst the reddest for their 
%metallicity. 

In order to quantify 
%place this result on a firmer foundation, we seek to
%quantify the metallicities and HB morphologies of each cluster. 
the metallicities,
%In the case of the metallicity measurement, 
we rely on the slope of the RGB.
%as an indicator of metallicity. 
Mighell et al. (\cite{migea1998}) have shown that metallicities
derived from the RGB slope are minimally affected by the age of the 
cluster. The measurement of the slopes and the transformation to
metallicities on the Zinn \& West (1984) system are both
described in Sarajedini et al. (\cite{sarea1998}).
Table 1 lists the measured metallicities for our program clusters.
Because of the extreme differential reddening in the region of
M33-R14, its abundance has been visually estimated via a direct comparison with
the RGBs of Da Costa \& Armandroff (\cite{daar1990}),
%instead of the RGB slope technique. 
Therefore, the estimated metallicity of R14 is rather uncertain.
Table 1 also lists the cluster metallicities from 
Christian \& Schommer (\cite{chsc1988}; $[Fe/H]_{CS}$) and 
Brodie \& Huchra (\cite{brhu1991}; $[Fe/H]_{BH}$), both of which have been
derived using integrated spectra. 
%The mean difference between
%these abundance estimates and those determined via the CMDs 
%(i.e. $[Fe/H]_{CMD} - [Fe/H]_{spectra}$) is 
%$0.23 \pm 0.25$ (s.e.m.) and $-0.22 \pm 0.21$ (s.e.m.), respectively.
%Neither of these offsets is statistically significant and both are
%consistent with our mean metallicity error of $\pm0.24$ dex.
Our values are in general intermediate between these other studies,
but the differences are not statistically significant. 
%We can also estimate the cluster metallicities
%using the integrated $B-V$ colors from Christian \& Schommer (\cite{chsc1988})
%along with our estimated reddenings and the equations of Couture et al. 
%(\cite{coutea1990}). This procedure yields a mean difference of
%$\langle$$[Fe/H]_{CMD} - [Fe/H]_{B-V}$$\rangle$$ = -0.12 \pm 0.20$ (s.e.),
%which provides further support for our abundance measures and their errors.
%All of these comparison calculations have been performed using the
%small-number statistical formulae of Keeping (\cite{keep1962}).
The reddening for each cluster (Table 1) follows from Eq. 1 of 
Sarajedini (\cite{sar1994}). The errors in the reddenings do not include 
the uncertainty in the photometric zeropoint. The significance of these 
values is discussed in Sarajedini et al. (1998).
%Excluding H10, which has a significantly higher reddening,
%we find $\langle$$E(V-I)$$\rangle$$ = 0.06 \pm 0.02$. This is in agreement with
%the maps of Burstein \& Heiles (\cite{buhe1982}), but differs from
%the study of Johnson \& Joner (\cite{jojo1987}), which concludes that
%$E(B-V)>0.077$ ($E(V-I) > 0.10$) in the direction of M33. However,
%the significant errors associated with our photometric zeropoint preclude
%us from making a stronger statement.

To quantify the HB morphologies, we rely upon generalized color histograms
of the HB stars. For the red HB clusters where all of the HB stars 
lie in a clump on or near the RGB, we adopt 
the peak of this histogram, i.e. $\sim$$(V-I)_g$, as the mean HB color. 
For the two clusters with significant blue
HB populations, we perform an `intensity-weighted' average of the
colors of the red and blue peaks.
%In the case of the M33 clusters, it is difficult to quantify the HB morphology
%using traditional techniques which rely on counting the numbers of stars 
%along the HB. The primary reason for this is that the 
%significant field contamination in each CMD adds an 
%uncertainty to the counts of HB stars that is difficult to quantify.
%However, in all of the clusters except U77 and M9, essentially 
%all of the HB stars are redward of the instability strip and lie in a clump 
%on or near the RGB. As a result, we use generalized histograms of the HB 
%stars to estimate the mean HB color.  
%mean dereddened color of the HB as an indicator
%of its morphology. 
%Our derived values are given in Table 1.
%First, we need to measure the V magnitude of the HB
%[V(HB)]. This was done by constructing a LF along the RGB and fitting a 
%Gaussian curve to the maximum in this LF. Then, limiting ourselves to 
%stars with magnitudes in the
%range $\pm0.3$ mag of $V(HB)$, we construct generalized color histograms
%of the HB stars. For the red clump clusters, we adopt the peak of this
%histogram, i.e. $\sim$$(V-I)_g$, as the mean HB color. For the two 
%clusters with significant blue
%HB populations, we perform an `intensity-weighted' average of the
%colors of the red and blue peaks. 
We therefore
have quantitative measures of each cluster's metallicity and intrinsic
mean HB color (i.e. HB morphology).

\section{Discussion}

To facilitate the comparison of these colors with those
of Galactic globular clusters and keeping in mind that the absolute errors in
our photometry are fairly significant, we will examine the difference
in color between the HB and RGB ($d_{B-V} = (B-V)_g - (B-V)_{HB}$). 
Although it makes very 
little practical difference, for the sake of consistency, we have 
converted the $(V-I)$ values of the M33 clusters to $(B-V)$ using a 
combination of the relation published by 
Zinn \& Barnes (\cite{ziba1996}) and one described by 
Sarajedini et al. (\cite{sarea1998}). 
%It is important to keep in mind
%that the uncertainties
%in these transformed colors are rather significant amounting to
%approximately $\pm0.05$ mag for clusters with $[Fe/H]<-1.3$ to
%as much as $\pm0.08$ mag for more metal-rich clusters.
The filled symbols in Fig. 2 are the observed values of $[Fe/H]$ and 
$d_{B-V}$ for Galactic globular clusters from 
Buonanno et al. (\cite{buoea1997}) and Sarajedini et al. (\cite{sarea1995}). 
In the former case, we have adopted the relation between metallicity and 
dereddened RGB color from Sarajedini \& Layden (\cite{sala1997}) to convert 
$\langle(B-V)_{o,HB}\rangle$ to $d_{B-V}$. The plus signs in Fig. 2 are 
7 `old' globular clusters in the SMC
from the work of Mighell et al. (\cite{migea1998}). The open circles show the
locations of the 9 M33 clusters for which we have reliable measurements. 
The values for M9 and U77 (clusters with a blue component to
their HBs) are indicated as lower limits because artificial
star experiments (described in Sarajedini et al. \cite{sarea1998})
indicate that our observations/reductions are not able to detect an extended
blue component to the HBs of these clusters, if in reality they possess
such a feature. In addition, some of the $d_{B-V}$ values for the red HB
clusters are slightly negative; this is simply due to measurement
errors, which could be as much as $\sim$0.05 mag. The dashed lines in Fig. 2 represent
the location of scaled-solar abundance Zero Age Horizontal Branches 
for masses of 0.62, 0.66, 0.74, and 0.90 $M_{\odot}$ from the work
of Dorman (\cite{dor1992}), again converted to $d_{B-V}$ via the relation
published by Sarajedini \& Layden (\cite{sala1997}).

Figure 2 shows that the general trend is for more metal-poor clusters to
have bluer HB colors, and this is corroborated by the theoretical
models. However, there is a significant dispersion in $d_{B-V}$ at a given
metallicity, especially at intermediate abundances. This is the so-called
``second parameter effect'' wherein another parameter, in addition to
metallicity, is influencing the morphology and thus the color of the HB.
Figure 2 suggests that, whatever this second parameter actually is, the
majority of halo globular clusters in M33 are significantly more affected 
by it than those in the MW or the SMC. In fact, the M33 clusters U49 and M9
are a second-parameter pair (similar to M3 and M13) in which both clusters
have similar metallicities, but vastly different HB morphologies.
%The recent review paper by Sarajedini et al. (\cite{sarea1997}) presents an
%in-depth discussion of the second parameter effect including references 
%to previous work on this topic. 
Among the many possible candidates for
the second parameter, Sarajedini et al. (\cite{sarea1997}) assert that
cluster age is the most likely one (although see 
Stetson et al. \cite{stvabo1996} for an
alternative viewpoint). If we adopt this assertion, then the
M33 globular clusters appear to be significantly younger, in the mean, than
similar clusters in the MW. 

There is at least one additional 
piece of evidence that supports the possibility of young
age for these M33 globular clusters. Several of them appear to
have a small number of stars located brighter than the first ascent RGB tip.
Figure 3 illustrates the CMDs of C20, H38, H10, and R12.
%these are the same datasubsets used in the determination of the clusters' 
%metallicities (see Sec. 3). 
The photometry has been shifted in color using the reddenings given in
Table 1 and in magnitude by requiring that $(V-I)_{o,-3}$ (the dereddened
color at $M_I = -3$) be equal to
that predicted by Equation 1 of Da Costa \& Armandroff (\cite{daar1990})
given the metallicity in Table 1. This is done instead of utilizing the 
RR Lyrae magnitude because these clusters have red HB clumps, which 
are generally 
not at the RR Lyrae level. The solid lines in Fig. 3 are the standard
RGB sequences of Da Costa \& Armandroff (\cite{daar1990}). 
The presence of stars above the 
magnitude of the first ascent red giant branch tip ($M_I$$\sim$--4.0)
for clusters with $[Fe/H]\lea-1.0$ supports 
the assertion that some of the M33 halo globular clusters are of young to 
intermediate age (see Mould \& Da Costa \cite{moda1988}).

We note that there were previous indications that {\it perhaps} some of these 
clusters may not be as old as globular clusters in our Galaxy 
(Cohen et al. \cite{copese1984}; Searle et al. \cite{sewiba1980}; 
Christian \& Schommer \cite{chsc1983}).
%Cohen et al. (\cite{copese1984}) derived SWB 
%(Searle et al. \cite{sewiba1980}) types for four clusters
%in our sample, and although their classifications of U49 and M9 as type VII 
%suggests old age (and indeed M9 at least appears to be a genuine old cluster), 
%they found C20 and H38 to be type VI clusters and therefore not as old as
%the oldest MW globulars. Also, the ages derived by Christian \& 
%Schommer (\cite{chsc1983}; \cite{chsc1988}) from integrated spectra and 
%colors only state that the age $\geq$4 Gyr for R12, R14, M9, H38, C20 and C38, 
%and that U49 is 
%only $\sim 2.5$ Gyr old, which is certainly in qualitative agreement with 
%our findings.
To place a lower limit on the ages of these clusters, we have compared the
CMD of C20 with the theoretical isochrones of 
Bertelli et al. (\cite{berea1994}). This comparison indicates that
the ages of the youngest M33 clusters must be older than $\sim$2 Gyr;
otherwise the main-sequence turnoffs would be bright enough to be detected
in our CMDs. Given this fact, we use the nomenclature of Hodge 
(\cite{hod1988}) to remind ourselves that ``globular" clusters
should be characterized by their mass ($>10^4M_{\sun}$) and not by their age.
Estimating an {\it upper} limit to the ages of these clusters is less
straightforward. Certainly, we can say that the two clusters with blue 
HB components (M9 and U77) are likely to be similar in age to the 
`younger-halo' MW globulars (such as M3). However, an upper limit
to the ages of the remaining
eight M33 globular clusters, which have red clumps located essentially
on top of their RGBs, is more difficult to ascertain.
From these observations, we can conclude that the halo kinematics 
reported by Schommer et al (\cite{schea1991}) for
this sample of objects implies that some intermediate-aged
M33 clusters are NOT in the main disk of that galaxy, which has
implications for cluster formation in a wide range of galaxies.
We also note in passing that, since only 20\% of our M33 clusters are
{\it bona fide} old globular clusters, the specific frequency of
such clusters in M33, which Schommer et al. (\cite{schea1991}) and 
Bothun (\cite{both1992}) have claimed is anomalously high ($\sim$60-70), may be 
reduced by a factor of $\sim$5.

\section{Implications for the Formation of the Local Group}

%We seek to place M33 into the formation history of the Local
%Group by comparing the M33 globular
%clusters with those in other Local Group galaxies.
%First, we note that b
By studying their main-sequence turnoffs, it has become 
clear that the oldest globular clusters in
the LMC are similar in age, within $\sim$1 Gyr, to those in the MW 
(Olsen et al. \cite{olea1998}). In addition, 
Mighell et al. (\cite{migea1998}) have used main-sequence turnoff 
photometry of the oldest globular clusters in the SMC and find 
that these are 3-4 Gyr younger than the oldest MW clusters.
The old clusters in the Sagittarius dwarf galaxy (Sgr), M54 and Terzan 8, 
have been studied by Layden \& Sarajedini (\cite{lasa1997}) and 
Montegriffo et al. (\cite{montea1998}), respectively. Both of these
clusters appear to be as old as MW globular clusters at their
metallicity.

We can correlate the findings based on main-sequence turnoff photometry
with the observed HB morphologies of the oldest clusters in each galaxy
discussed above. For example, the oldest globular clusters in the MW,
LMC, and Sgr all possess HBs that are populated predominantly blueward of 
the RR Lyrae
instability strip. In contrast, the oldest clusters in the SMC, which are
3-4 Gyr younger than those in the MW, LMC, and Sgr, have
predominantly red HB morphologies. This trend of redder HB morphology
with younger age is, in our opinion, simply the second parameter effect noted above.

%We can now utilize the HB types of globular clusters to place M31 and M33 
%in this global chronology. 
A number of studies have exploited the superior resolution of HST to
construct CMDs for M31 globulars (Ajhar et al. \cite{ajhea1996};
Fusi Pecci et al. \cite{fusea1996}; Holland et al. (\cite{holea1997})).
The collective results of these investigations reveal the existence
of metal-poor blue HB clusters, which seems to indicate that the oldest
globular clusters in M31 are probably similar in age to those in the 
MW (see also Djorgovski et al. \cite{djorea1997}). 
Along the same lines, in the halo of M33, we have uncovered two 
intermediate-to-metal-poor clusters with blue+red HBs. Thus, based 
upon the appearance of 
Fig. 2 and the implications of theoretical HB models 
(e.g. see Lee et al. {\cite{ldz1994}), we can be reasonably certain 
that, even though the MW, M31, M33, LMC, and Sgr are spread across a
region of space several hundred kiloparsecs in size, they all formed 
their first generation 
of clusters at around the same time soon after the Big Bang
(see also Harris et al. 1997).  Ages
determined from cluster main-sequence turnoffs indicate that 3 to 4 Gyr
later, the first globular in the SMC (NGC 121) formed. 
Lastly, the bulk of the M33 clusters (those with purely 
red HBs) followed several Gyrs later still. If we consider only the galaxies
with `halo' populations, we find that the age range among the halo
clusters in M33 is somewhere between a factor of 2 to 3 larger than those of the
MW and M31. 

% The \notetoeditor{TEXT} command allows the author to communicate some
% information to the copy editor.  This information will appear as a 
% footnote on the printed copy for the aasms4 style file.  Nothing will 
% appear on the printed copy if the aaspp4 or aas2pp4 style file is used.

% In these sections, we see some additional math-related markup, and we
% have references to one of the tables (occurs later in the document)
% and the "hairymath" section immediately preceding this one.
%
% In the second paragraph, note the use of in-text math ($stuff$) including
% a couple of the miscellaneous symbol commands defined in the AASTeX macro
% package.
%
% This is the last section of the paper, so there is an \acknowledgments
% section at the end of the main body.

\acknowledgments

It is a pleasure to acknowledge P. B. Stetson for kindly providing the 
ALLFRAME software and WFPC2 PSFs for our use.
We are grateful to Alistair Walker, Chip Kobulnicky, and Eva Grebel
for providing valuable comments on an earlier version of this manuscript.
This research was supported by the National Aeronautics and Space 
Administration (NASA) grant numbers HF-01077.01-94A and
GO-05914.01-94A (AS), GO-05914.02-94A (DG), GO-05914.03-94A (RS),
GO-05914.04-94A (PH) from the Space Telescope Science
Institute, which is operated by the Association of Universities for
Research in Astronomy, Inc., under NASA contract NAS5-26555.
\newpage 
%
% Figure 1:
%
\figcaption[fig1.eps]{
\label{fig-1}
The top two panels show CMDs for M9 and C20, which
are two halo globular 
clusters in M33. The solid lines are the fiducial sequence of the Galactic
globular cluster M5 (Sandquist et al. 1996) offset to match the 
$V(HB)$ and $(V-I)_g$ of M9 and
C20. The lower panels show CMDs of the Galactic globular clusters 
M3, which is similar to M33-M9 in metal abundance, and NGC 362, which 
is similar to M33-C20 in metal abundance. The M33 CMDs are V vs. V--I while 
those for the Galactic globular clusters are V vs. B--V.}

%%
%% Figure 2: 
%%
\figcaption[fig2.eps]{
\label{fig-2}
The difference in color between the HB and RGB ($d_{B-V}$) 
for MW globular clusters (filled circles), SMC globular clusters 
(plus signs), and the
M33 halo globular clusters (open circles). The 
dashed lines are the expected locations of zero-age horizontal branch 
models for 0.62, 0.66, 0.74, and 0.90 $M_{\odot}$ (from right to left) from the 
work of Dorman (1992).}

%%
%% Figure 3: 
%%
\figcaption[fig3.eps]{
\label{fig-3}
Color-magnitude diagrams for four of the M33 clusters in our sample. The
solid lines are the standard red giant branches of M15 ($[Fe/H] = -2.2$),
NGC 1851 ($[Fe/H] = -1.3$), and 47 Tuc ($[Fe/H] = -0.7$) from Da Costa \&
Armandroff (1990). 
%Note the existence of cluster stars brighter than
%the RGB tip magnitude; this is consistent with a young to intermediate
%age for these clusters.
}

\newpage
\begin{deluxetable}{lcccccccc}
\scriptsize
\tablewidth{0in}
\tablenum{1}
\tablecaption{M33 Globular Cluster Characteristics}
\tablehead{
\colhead{Cluster} &
\colhead{$V$}  &
\colhead{$B-V$}  &
\colhead{$R$\tablenotemark{a}}  &
\colhead{$[Fe/H]_{CS}$\tablenotemark{b}} &
\colhead{$[Fe/H]_{BH}$\tablenotemark{c}} &
\colhead{$[Fe/H]_{CMD}$\tablenotemark{d}} &
\colhead{$E(V-I)$} &
\colhead{HB Morphology} }
\startdata
 U49 & 16.25 & 0.68 & 8.4 & $-0.8 \pm 0.3$ & $-1.70 \pm 0.53$ & $-1.64 \pm 0.20$
& $0.07 \pm 0.02$ & Red \cr
 R12 & 16.38 & 1.03 & 4.1 & $-1.2 \pm 0.3$ & \nodata          & $-1.19 \pm 0.24$
& $0.05 \pm 0.03$ & Red \cr
 R14 & 16.48 & 0.98 & 2.5 & $-1.5 \pm 0.3$ & \nodata          & $-1.0 \pm 0.5$  &
 \nodata         & Red \cr
 M9  & 17.12 & 0.69 & 8.9 & $-1.7 \pm 0.3$ & \nodata          & $-1.64 \pm 0.28$
& $0.04 \pm 0.03$ & Red+Blue \cr
 U77 & 17.19 & 0.67 & 5.4 & \nodata        & $-1.77 \pm 0.77$ & $-1.56 \pm 0.30$
& $0.08 \pm 0.03$ & Red+Blue \cr
 H38 & 17.25 & 0.73 & 11.1& $-1.5 \pm 0.3$ & \nodata          & $-1.10 \pm 0.10$
& $0.04 \pm 0.01$ & Red \cr
 C20 & 17.67 & 0.77 & 15.3& $-2.2 \pm 0.3$ & $-1.25 \pm 0.79$ & $-1.25 \pm 0.22$
& $0.03 \pm 0.03$ & Red \cr
 C38 & 18.10 & 0.73 & 17.5 & $-1.2 \pm 0.3$ & \nodata         & $-0.65 \pm 0.16$
& $0.04 \pm 0.02$ & Red \cr
 H10 & 18.23 & 0.96 & 10.5& \nodata        & $-0.91 \pm 0.90$ & $-1.44 \pm 0.26$
& $0.25 \pm 0.03$ & Red \cr
 U137 & 18.30 & 0.83 & 12.8 & \nodata      & $-0.12 \pm 0.38$ & $-0.98 \pm 0.16$
& $0.09 \pm 0.03$ & Red \cr
\enddata
\tablenotetext{a}{Radial distance in arcminutes from the center of M33.}
\tablenotetext{b}{Christian \& Schommer (1988)}
\tablenotetext{c}{Brodie \& Huchra (1991)}
\tablenotetext{d}{Metallicity based on the color-magnitude diagram.}
\end{deluxetable}
%%
%%=============================================================================

\begin{figure}
\epsfxsize=7in \epsffile{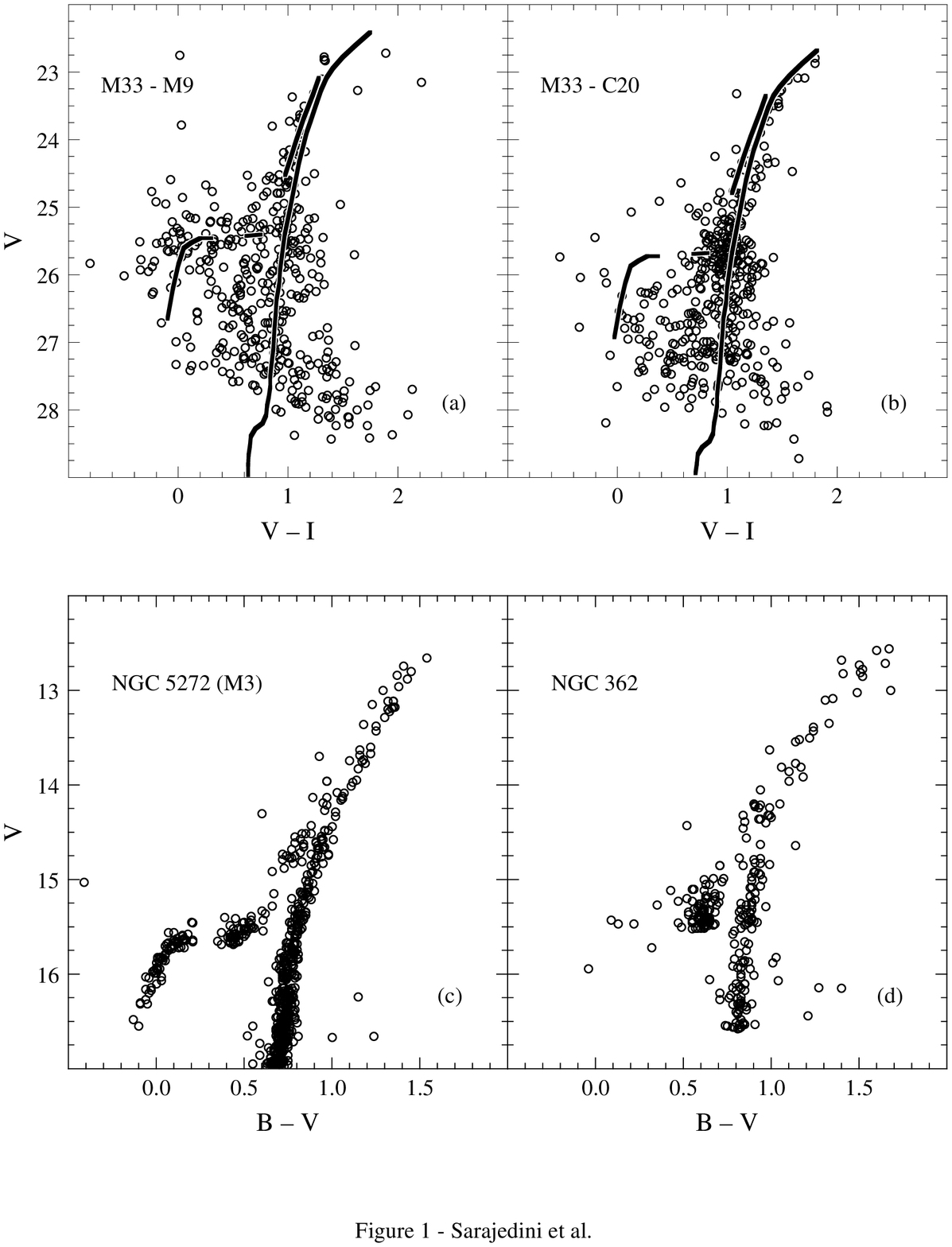}
\end{figure}

\begin{figure}
\epsfxsize=7in \epsffile{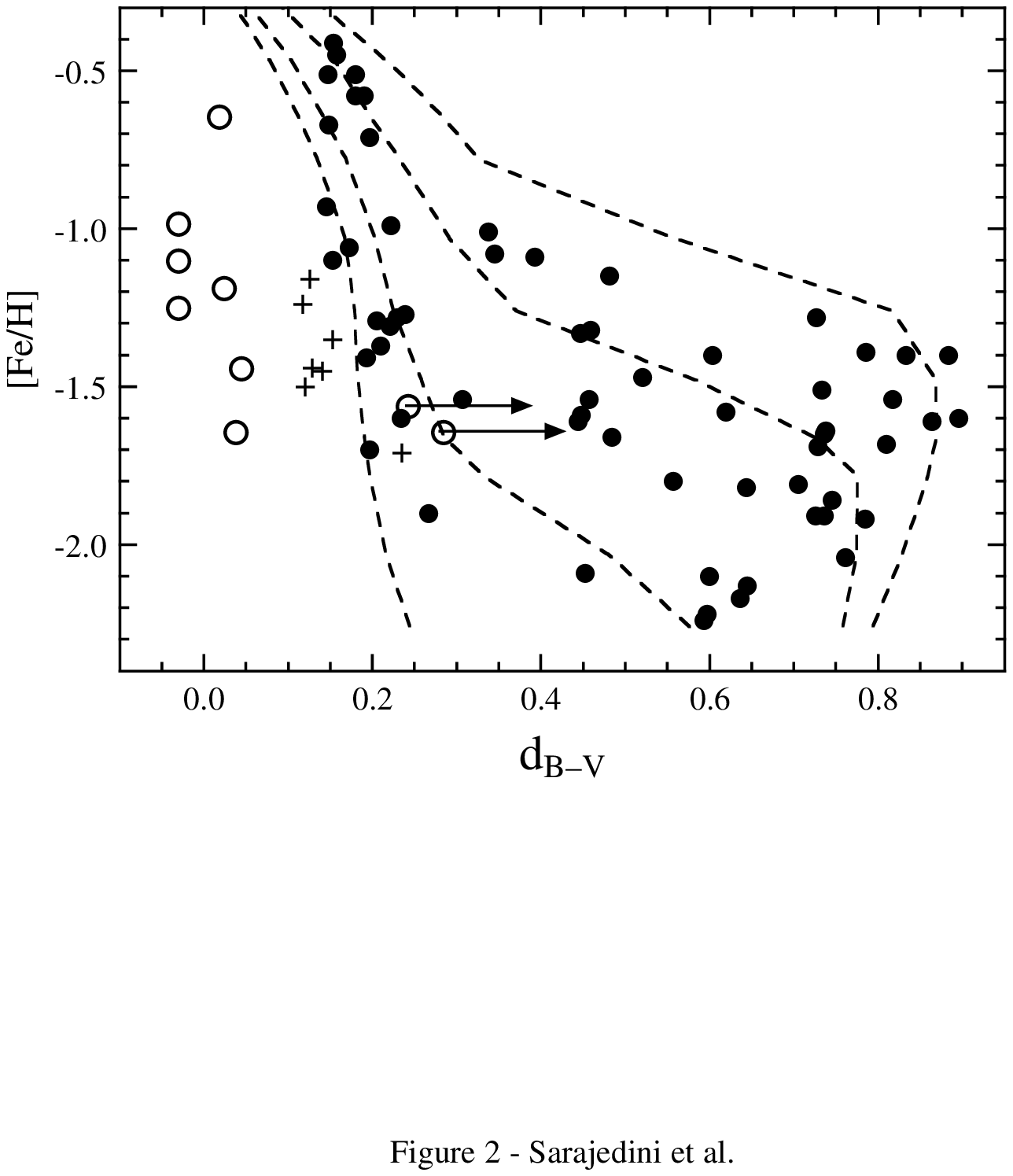}
\end{figure}

\begin{figure}
\epsfxsize=7in \epsffile{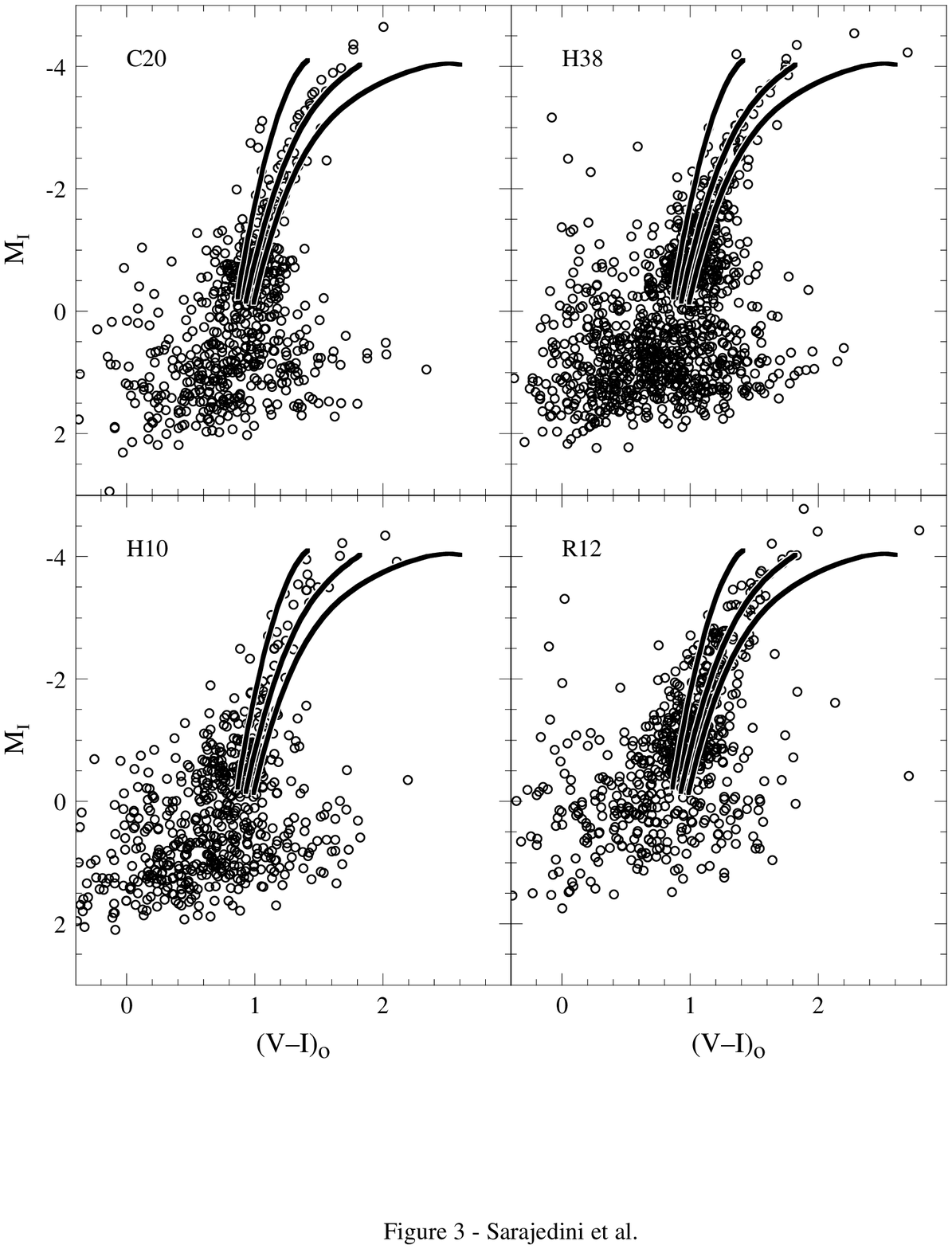}
\end{figure}

\end{document}